\documentclass[a4paper,fleqn]{cas-sc}

\usepackage[numbers]{natbib}

\usepackage{amssymb}
\usepackage{graphicx}

\usepackage{xcolor}
\usepackage{colortbl}
\definecolor{shadecolor}{gray}{0.95}

\usepackage{tcolorbox}

\usepackage{multirow}
\usepackage{multicol}

\usepackage{booktabs}

\usepackage{arydshln}

\usepackage{caption}
\usepackage{subcaption}

\usepackage{verbatim}
\usepackage{enumitem}
\usepackage{soul}

\definecolor{lightblue}{rgb}{0.13, 0.59, 0.82}

\definecolor{myPurple}{RGB}{128,0,128}
\definecolor{myBlue}{RGB}{0,176,240}
\definecolor{myGreen}{RGB}{0,176,80}
\definecolor{myRed}{RGB}{255,0,0}
\definecolor{darkgreen}{rgb}{0,0.5,0}
\definecolor{myGray}{RGB}{175,171,171}
\definecolor{myOrange}{RGB}{255,192,0}

\begin{document}
\let\WriteBookmarks\relax
\def\floatpagepagefraction{1}
\def\textpagefraction{.001}

\shorttitle{E-code: Mastering Efficient Code Generation through Pretrained Models and Expert Encoder Group}    

\shortauthors{Yue Pan, Chen Lyu, Zhenyu Yang, Lantian Li, Qi Liu, Xiuting Shao }

\title [mode = title]{E-code: Mastering Efficient Code Generation through Pretrained Models and Expert Encoder Group}

% % Title footnote mark
% % eg: \tnotemark[1]
% \tnotemark[<tnote number>] 

% % Title footnote 1.
% % eg: \tnotetext[1]{Title footnote text}


\author[SDNU2,SDNU1]{Yue Pan}           \ead{pany@mail.sdu.edu.cn}
\author[SDNU1]{Chen Lyu}\corref{cor1}   \ead{lvchen@sdnu.edu.cn}
\author[SDNU2]{Zhenyu Yang}             \ead{yangzycs@mail.sdu.edu.cn}
\author[SDNU2]{Lantian Li}              \ead{lilantian@mail.sdu.edu.cn}
\author[SDNU1]{Qi Liu}                  \ead{1642339035@qq.com}
\author[SDNU1]{Xiuting Shao}            \ead{shaoxiuting@126.com}

\cortext[cor1]{Corresponding author}

\affiliation[SDNU2]{organization={Shandong University},
            % addressline={Shandong Normal University}, 
             city={Qingdao},
% %          citysep={}, % Uncomment if no comma needed between city and postcode
%             # postcode={250301}, 
            state={Shandong},
            country={China}}
            
\affiliation[SDNU1]{organization={School of Information Science and Engineering, Shandong Normal University},
            % addressline={Shandong Normal University}, 
            city={Jinan},
%          citysep={}, % Uncomment if no comma needed between city and postcode
            postcode={250301}, 
            state={Shandong},
            country={China}}

% % Corresponding author indication
% \cormark[<corr mark no>]

% % Footnote of the first author
% \fnmark[<footnote mark no>]

% % Email id of the first author
% \ead{<email address>}

% % URL of the first author
% \ead[url]{<URL>}

% % Credit authorship
% % eg: \credit{Conceptualization of this study, Methodology, Software}
% \credit{<Credit authorship details>}

% % Address/affiliation
% \affiliation[<aff no>]{organization={},
% %          citysep={}, % Uncomment if no comma needed between city and postcode

% % Footnote of the second author

% % Email id of the second author
% \ead{}

% % URL of the second author

% % Credit authorship

% % Address/affiliation
% \affiliation[<aff no>]{organization={},
% %          citysep={}, % Uncomment if no comma needed between city and postcode

% % Corresponding author text
% \cortext[1]{Corresponding author}

% % Footnote text

% % For a title note without a number/mark
% %\nonumnote{}

% \begin{abstract}

% \end{abstract}
\begin{abstract}
\noindent\textbf{Context:} With the waning of Moore's Law, the software industry is placing increasing importance on finding alternative solutions for continuous performance enhancement. The significance and research results of software performance optimization have been on the rise in recent years, especially with the advancement propelled by \textbf{L}arge \textbf{L}anguage \textbf{M}odel\textbf{s} (LLMs). However, traditional strategies for rectifying performance flaws have shown significant limitations at the competitive code efficiency optimization level, and research on this topic is surprisingly scarce.

\noindent\textbf{Objective:} This study aims to address the research gap in this domain, offering practical solutions to the various challenges encountered. Specifically, we have overcome the constraints of traditional performance error rectification strategies and developed a \textbf{L}anguage \textbf{M}odel (LM) tailored for the competitive code efficiency optimization realm.

\noindent\textbf{Method:} We introduced E-code, an advanced program synthesis LM. Inspired by the recent success of expert LMs, we designed an innovative structure called the Expert Encoder Group. This structure employs multiple expert encoders to extract features tailored for different input types. We assessed the performance of E-code against other leading models on a competitive dataset and conducted in-depth ablation experiments.

\noindent\textbf{Results:} Upon systematic evaluation, E-code achieved a 54.98\% improvement in code efficiency, significantly outperforming other advanced models. In the ablation experiments, we further validated the significance of the expert encoder group and other components within E-code.

\noindent\textbf{Conclusion:} The research findings indicate that the expert encoder group can effectively handle various inputs in efficiency optimization tasks, significantly enhancing the model's performance. In summary, this study paves new avenues for developing systems and methods to assist programmers in writing efficient code.
\end{abstract}


% Keywords

\maketitle

% % Main text
% \section{}\label{}
\section{Introduction}\label{sec:1}

In recent years, with the proliferation of \textbf{computation-intensive applications}, the \textbf{execution efficiency} of software has garnered widespread attention. On one hand, Sam Altman's proposition that ``intelligence in the universe doubles every 18 months'' suggests that our computational demand will continue to grow exponentially in the future. On the other hand, the automatic performance improvements brought about by Moore's Law are gradually slowing down \cite{patterson2021carbon}. Given this context, the execution efficiency of software has become a pivotal factor determining the success of software applications.

\textbf{Performance optimization} has emerged as a \textbf{pressing challenge}. Unlike functional defects, performance defects do not directly lead to system crashes. However, their adverse effects on user experience, system throughput, and resource utilization cannot be underestimated. It is particularly noteworthy that, compared to functional defects, performance defects are often harder to detect \cite{attariyan2012x,dean2014perfscope,catchmeifyoucan} and rectify \cite{nistor2013,song2014oopsla}. This not only increases the burden on developers but also underscores the inadequacies of current tools in pinpointing and addressing such issues. For novice developers, this is undoubtedly a monumental challenge, as they might lack the requisite knowledge and experience to tackle such subtle yet critical performance issues. Hence, there's usually a reliance on experts in the domain of performance optimization. These experts are capable of taking into account various factors such as the internal logic of the code, functional requirements, and limitations of running time environments to offer the best optimization strategies  \cite{chen2019inferring}.

However, the \textbf{current methods} for rectifying performance defects still have evident  \textbf{limitations}. While numerous performance defect rectification methods based on specific algorithms and rule sets exist, most of these methods only target certain specific problems, such as redundant calculations \cite{memoization}, software configuration errors \cite{misconfigurations}, or inefficient loops \cite{nistor2013,linhai2017icseloops,xiao2013issta}. This specificity makes the rectification of performance defects more intricate and also makes the construction and maintenance of these rule sets both time-consuming and resource-intensive \cite{bielik2017learning}.

\textbf{L}arge \textbf{L}anguage \textbf{M}odel\textbf{s} (LLMs) have opened up a \textbf{new research} direction for \textbf{performance optimization}. In the field of \textbf{N}atural \textbf{L}anguage \textbf{P}rocessing (NLP), LLMs have achieved significant breakthroughs \cite{brown2020language} and are gradually showcasing their potential in software engineering tasks such as code completion \cite{svyatkovskiy2020intellicode}, automatic documentation generation  \cite{clement2020pymt5}, unit test generation \cite{Tufano2020UnitTC}, and defect detection \cite{Drain2021DeepDebugFP}. Notably, LLMs possess the potential to learn and deduce performance optimization patterns from vast amounts of data, providing a fresh perspective and approach for software performance optimization. Currently, models like ChatGPT\footnote{\url{https://chat.openai.com}} \cite{chatgpt} have already shown promise in assisting novice programmers to some extent.

However, current LLMs still have certain \textbf{shortcomings}. For tasks related to code performance optimization, these models face significant challenges. Such tasks require the models to deeply comprehend complex \textbf{N}atural \textbf{L}anguage (NL) descriptions and master numerous algorithms and optimization strategies, a difficulty level that is formidable even for experienced programmers. It's worth noting that existing LLMs, such as CodeX \cite{chen2021evaluating} and AlphaCode \cite{li2022competition}, often exhibit what is termed the ``slow positive'' issue in the code they generate. This means that while the generated code functionally meets requirements, it doesn't achieve the anticipated performance standards. This phenomenon underscores not only the importance of code performance optimization but also indicates that current LLMs still need refinement in their knowledge of performance optimization. Moreover, the efficiency of code is often relative; the efficiency of one code segment can only be ascertained when compared to others. To grasp this differential knowledge, the model might need to receive multiple types of input simultaneously, such as NL descriptions of code functionality, efficient code examples, and inefficient code examples. Regrettably, existing LLMs are seldom designed to handle tasks with such varied inputs. When attempting to process multiple inputs, they often simply concatenate various inputs, which can lead to exceeding token length limitations, and might even result in adverse task transfer effects. For instance, when the input comprises both NL and code, to reduce interference from NL features when the model learns code, code comment markers are often added before the NL description.

Based on the above analysis, generating efficient code requires overcoming \textbf{numerous challenges}. To maximize the potential of LLMs in generating efficient code and to steer code models towards precise predictions, we have identified two key challenges: 1) Flexibility of code efficiency optimization strategies, and 2) Simultaneous extraction of features from different types of inputs.

\textbf{- Challenge 1: Flexibility of code efficiency optimization strategies.} Traditional performance defect rectification strategies tend to target only localized code modifications. Such optimization strategies are often confined to minor code adjustments or the inclusion of decorative statements. A fundamental reason for this is that these strategies, when guiding model optimization, primarily provide inefficient code as references without accompanying NL descriptions of the code's functionality. This leads to relatively constrained optimization strategies for the model. Therefore, in certain scenarios, such rectification strategies may manifest significant inflexibility.

\textbf{- Challenge 2: Simultaneous extraction of features from different types of inputs.} Conventional \textbf{L}anguage \textbf{M}odels (LMs), when receiving inputs, tend to extract their features holistically, which is undoubtedly efficient when dealing with indivisible inputs. However, in many real-world scenarios, models might need to handle multiple inputs of different types or meanings and extract features from these inputs collectively. For instance, when learning knowledge about code efficiency, a model must process multiple types of inputs simultaneously, such as NL descriptions, inefficient code, and efficient code.

To address \textbf{Challenge 1}, we explored performance optimization strategies based on code generation. We introduced an entirely new ``\textbf{E}fficient \textbf{P}rogramming \textbf{C}ompetition'' (abbreviated as EPC) tasks as an optimization framework. Within this framework, performance optimization is viewed as a code generation problem. Specifically, when the model receives inefficient source code and its associated NL description, its primary goal is to generate a completely optimized new code segment, rather than merely patching or locally adjusting the original code. For performance defects caused by unsuitable data structures or algorithms, simple local modifications may not fundamentally solve the issue. To substantiate this viewpoint, we further present two empirical case studies, demonstrating that in certain contexts, code generation-based optimization strategies are superior to traditional local modification strategies.

\begin{center}
\begin{tcolorbox}[colback=gray!5,%gray background % 背景 颜色
                  colframe=black,% black frame colour % 字体颜色
                  width=16cm,% Use 8cm total width,%文本框宽度
                  arc=2mm, auto outer arc, % 文本框弧度
                  boxrule=0.5pt, %
                  toprule=1.2pt, rightrule=1.2pt, bottomrule=1.2pt,leftrule=1.2pt, % 边框粗细
                  ]

\textbf{Sorting Operation.} Consider an algorithm whose task is to receive an array and rearrange it so that the two largest elements are at the front. An intuitive but inefficient solution is to directly use a sorting function, sorting the entire array. This method is based on the Timsort algorithm with a time complexity of O($n\log_{}{n}$). When we only observe this inefficient code, optimizing it becomes nearly impossible since it outputs a fully sorted array. However, if we have a NL description of the algorithm, we can consider the heap sort method, using the heapq function \cite{floyd1964algorithm}, which has a time complexity of only O($2\log_{}{n}$).

\end{tcolorbox}
\end{center}

\begin{center}
\begin{tcolorbox}[colback=gray!5,%gray background % 背景 颜色
                  colframe=black,% black frame colour % 字体颜色
                  width=16cm,% Use 8cm total width,%文本框宽度
                  arc=2mm, auto outer arc, % 文本框弧度
                  boxrule=0.5pt, %
                  toprule=1.2pt, rightrule=1.2pt, bottomrule=1.2pt,leftrule=1.2pt, % 边框粗细
                  ]

\textbf{Search Operations.} Imagine a system using a linear array structure for data search, with a time complexity of O($n$). Solely relying on local optimization, it's challenging to achieve fundamental performance improvement. A more efficient strategy is to adopt a hash table in place of the array, thereby reducing the search operation's time complexity to O($1$).

\end{tcolorbox}
\end{center}

The \textbf{code generation-based} method offers greater \textbf{flexibility}. The EPC task provides a fresh perspective on code optimization. It not only breaks the limitations of local code optimization but also presents a flexible and efficient platform. This strategy has the potential to address a series of performance bottlenecks that traditional performance rectification strategies struggle with, ensuring efficient code execution across various scenarios.

To tackle \textbf{Challenge 2}, we introduced an innovative encoding mechanism — the Expert Encoder Group. The core strategy of the expert encoder group is to deploy targeted expert encoders for efficient and precise feature extraction from various inputs. Jang et al. \cite{jang2023exploring} has shown that training individual expert language models for each specific task is superior to a single multi-task fine-tuning approach. Inspired by this, we proposed the structural design of the expert encoder group, where the model uses multiple encoders to process different types of inputs individually. Unlike traditional LMs that simply concatenate all inputs for holistic feature extraction, our expert encoder group extracts features separately for each type of input. For instance, for the EPC task, the model needs to simultaneously handle various inputs such as inefficient code, algorithm tags, problem descriptions, input formats, output formats, and I/O test samples. Thus, we deployed six specialized expert encoders for feature extraction.

In this study, we introduce a code generation model named \textbf{E-code}, specifically designed for the \textbf{EPC tasks}. Inspired by the success of distributed training for language models \cite{don2023cold,wortsman2022fi,li2022branch}, E-code is composed of multiple independent but integrated language models. The technical highlight of E-code lies in its adoption of an innovative encoding mechanism, realized based on our designed expert encoder group. Recognizing the limitations of existing evaluation metrics, we also trained a language model called \textbf{ExecTimePredictor} to accurately predict the execution time of the code. On competition-level datasets, E-code improved code performance by \textbf{30.74\%}, significantly surpassing other leading LMs. Subsequently, utilizing a filtering mechanism based on ExecTimePredictor, E-code's performance was further enhanced to \textbf{54.98\%}. Additionally, we delved into the impact of the complexity of NL descriptions on LMs performance. Finally, through a series of ablation experiments, we validated the efficacy and practicality of the expert encoder group and other design approaches proposed in this study.

Our contribution falls into the following  aspects:
\begin{itemize}
\item \textbf{Model.} We introduce \textbf{E-code}, an innovative program synthesis language model. E-code is the first language model that adopts a code generation strategy for performance optimization.
\item \textbf{Encoding Methods.} We designed the \textbf{Expert Encoder Group}, a novel encoder architecture. This architecture aims to perform specialized feature extraction for various information types present in the input.
\item \textbf{Metric.} We introduce the \textbf{ExecTimePredictor} and \textbf{CodeExecTimeDB}, a model for predicting code execution time and its training dataset, respectively.
\item \textbf{Experiments.} On \textbf{competition-level} datasets, the performance of the E-code model significantly surpassed that of other advanced models. Concurrently, through \textbf{ablation} experiments, we further confirmed the practicality and efficacy of the expert encoder group and our other design decisions.
\end{itemize}

\section{Related Work}\label{sec:2}

\subsection{Program Synthesis}\label{sec:2.1}

The objective of \textbf{program synthesis} is to \textbf{automatically generate program code} based on specified specifications or constraints. These specifications can be presented in various ways, such as NL descriptions, I/O example sets, or explicit mathematical constraints  \cite{summers1977methodology, gulwani2012spreadsheet}. To generate programs that meet these specifications, highly specialized optimization algorithms are typically required \cite{rajeev1904sygus, li2023understanding}. In this regard, insights from Gulwani et al. \cite{gulwani2017program} have been invaluable in exploring the adaptability and efficiency of program synthesis methods.

In recent years, \textbf{pre-trained LMs} based on the large Transformer architecture \cite{vaswani2017attention} have shown \textbf{significant capabilities} in the field of program synthesis \cite{chen2021evaluating,nijkamp2022conversational}. These models can not only understand and generate complex programs involving a variety of general programming languages \cite{yin-neubig-2017-syntactic, xu2018sqlnet, chen2021evaluating}, but their program specifications are often based on NL descriptions \cite{hendrycksapps2021, austin2021program, poesia2022synchromesh}, expanding their range of applications. This momentum has promoted the widespread adoption of pre-trained language models in program synthesis, as they can extract rich contextual information from extensive code repositories and NL resources \cite{feng-etal-2020-codebert, clement-etal-2020-pymt5, codet5, gpt-j, ChenVarCLR2022}.

Currently, LLMs have demonstrated \textbf{impressive code generation capabilities}. However, there's a relative \textbf{weakness} in their \textbf{knowledge} of \textbf{code efficiency}, creating a \textbf{stark contrast} between these two aspects. The CodeX system has been successfully applied to GitHub Copilot, and the code generation capability of AlphaCode \cite{li2022competition} can even rival human programmers. It is worth noting, however, that although these models excel in code generation, the efficiency of the code they produce often falls short, with common issues exemplified by the ``slow positive'' phenomenon \cite{lertbanjongngam2022empirical}. This strongly indicates that generating efficient, high-performance code remains a significant challenge in the current landscape.

\subsection{Performance Bug Detection \& Fix}\label{sec:2.2}

\textbf{Performance error detection} and \textbf{repair} is a \textbf{rich research domain}, encompassing a variety of tools and methods designed specifically to locate and rectify performance bottlenecks \cite{yang2024narrepair}. Most of these tools aim to identify high-latency execution segments in the code, with many being crafted especially for performance testing, capable of automatically generating or selecting appropriate test cases. In addition, a range of specific types of performance issues has garnered widespread attention \cite{zhang2011ASE, Grechanik2012ICSE, burnim2009ICSE}. For instance, there are tools dedicated to detecting running time bloat \cite{xu-pldi-2009, xu2010pldi1, dufour2008fse}, inefficient data structures \cite{xu2010pldi}, database performance anti-patterns \cite{chen2014icse}, improper resource sharing in multi-threaded software \cite{liu2011oopsal}, and inefficient loops \cite{nistor2013, xiao2013issta}. To fix these performance issues, researchers have developed several targeted solutions, such as reducing redundant computations \cite{memoization}, rectifying software configuration errors \cite{misconfigurations}, and enhancing loop efficiency \cite{caramelnistor}. The tool we propose employs an innovative optimization strategy to mitigate common performance bottlenecks. By integrating the characteristics of EPC tasks, we utilize code generation from scratch to optimize efficiency, paving new research directions in this domain.

\subsection{Distributed Training of Language Models}\label{sec:2.3}

Recently, \textbf{distributed training} has seen a \textbf{growing trend} in the application of language models. For instance, Li et al. \cite{li2022branch} found that by independently training sub-models on different training subsets and subsequently merging them, a model with lower perplexity can be obtained, significantly outperforming a one-time training on the entire dataset. Wortsman et al. \cite{wortsman2022model} explored how to enhance the fine-tuning efficiency of language models. They proposed a strategy to merge multiple models aimed at the same task but with different configurations to boost task performance and demonstrated that fine-tuning with different subsets can further enhance model efficiency. Don-Yehiya et al. \cite{don2023cold} discussed how to fine-tune language models on different tasks from a multi-task learning perspective and employed distributed methods for merging. They pointed out that not only can this approach construct multi-task fine-tuned models, but it also possesses other benefits, such as federated learning \cite{mcmahan2017communication}. In this paper, our introduced E-code and expert encoder group predominantly adopt the strategy of merging independently trained models to achieve both performance and efficiency enhancements. This method further corroborates the feasibility and advantages of distributed training in the realm of language models. 
In the field of code generation using deep learning, E-code is the first model specifically designed to focus on generating code performance.

\section{Motivation}\label{sec:3}

\subsection{Motivating Example}\label{sec:3.1}

\begin{figure}
    \centering
    \includegraphics[width=0.95\linewidth]{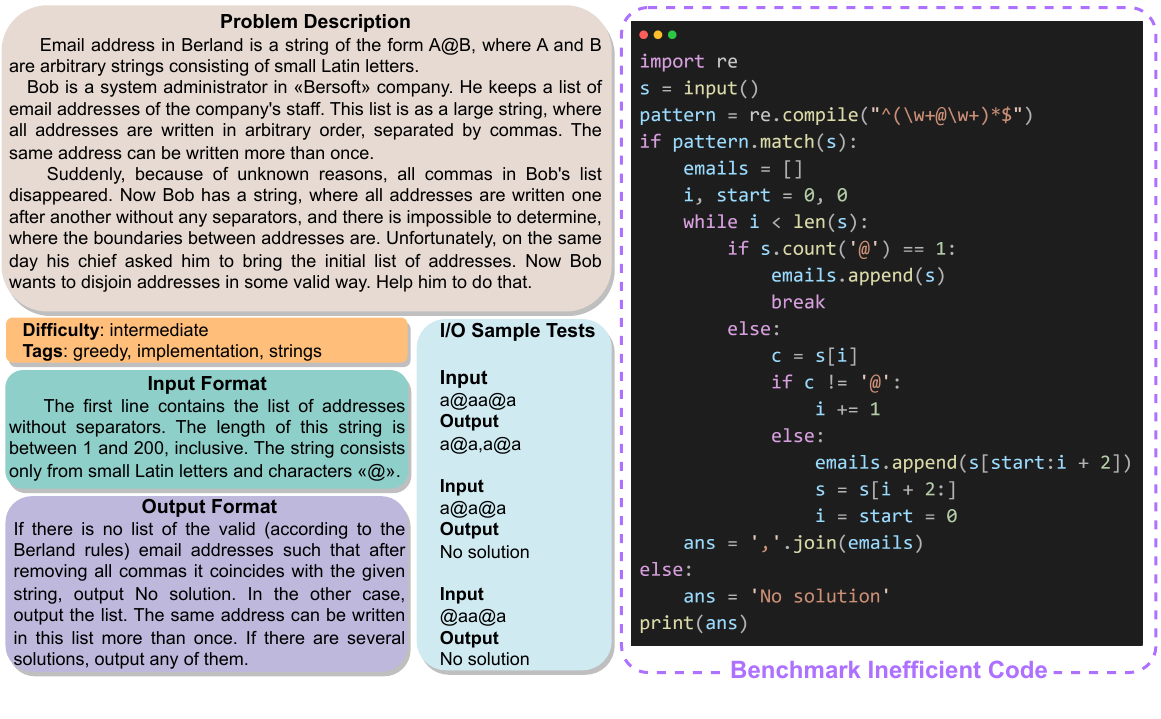}
    \caption{An NL description of a problem from the CodeForces programming competition website and its associated inefficient solution code (running time 266 µs).}
    \label{fig:Motivating Example}
\end{figure}

\begin{figure}
    \centering
    \includegraphics[width=0.95\linewidth]{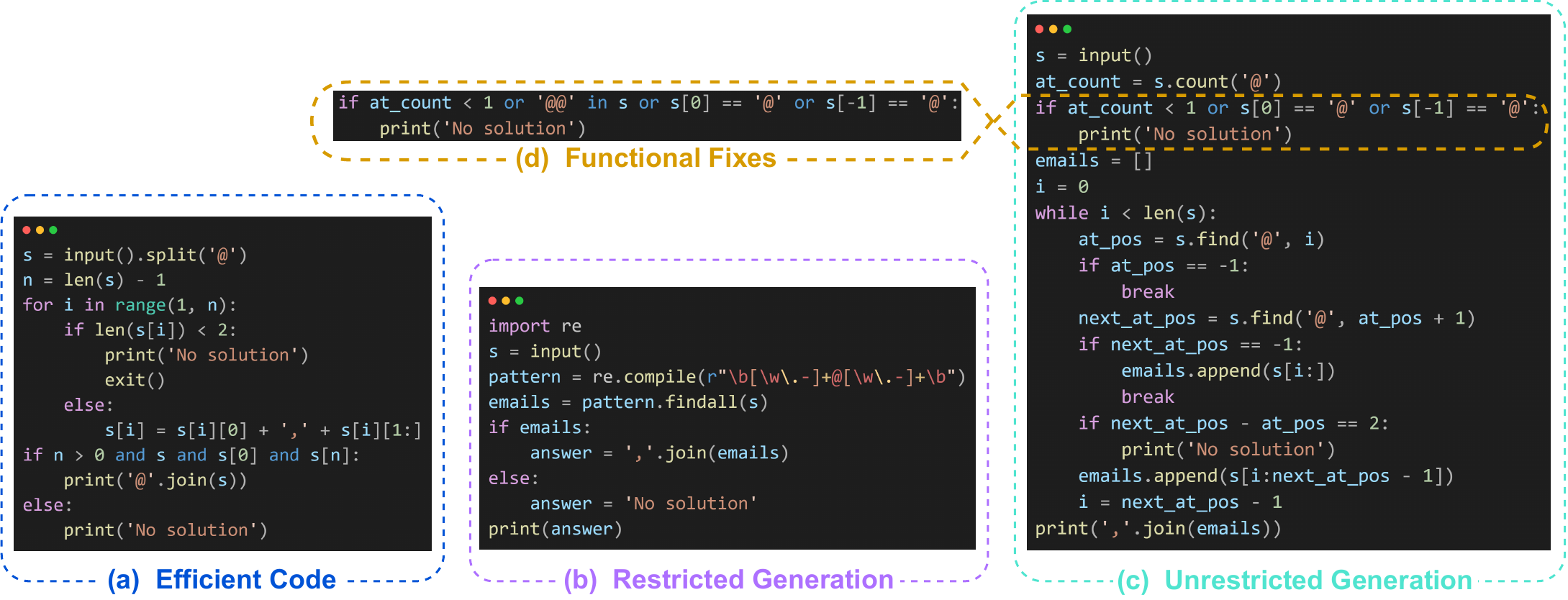}
    \caption{Displays the problem's related outcomes. This figure covers four aspects: 
    (a) the sample's efficient code (running time 22 µs); 
    (b) ChatGPT-4's code generation results solely based on inefficient code (running time 251 µs); 
    (c) ChatGPT-4's code generation results when exposed to all the information (running time 25 µs); and 
    (d) ChatGPT-4's code adjustments after being individually reminded about I/O tests (running time 25 µs).}
    \label{fig:Motivating Example Code}
\end{figure}

Figure \ref{fig:Motivating Example} displays an NL description of a problem\footnote{\url{https://codeforces.com/problemset/problem/31/B}} taken from the CodeForces\footnote{\url{https://codeforces.com}} programming competition website, along with its associated inefficient solution code. The problem requires the splitting of strings following the ``A@B'' pattern. The inefficient code (running time 266 µs) starts by constructing a regular expression to determine feasibility. Then, it carries out a string splitting operation, ensuring that every string adheres to the ``A@B'' pattern, and finally outputs the result connected by commas. However, a more efficient algorithmic (running time 22 µs) approach seems to employ a different method, as shown in Figure \ref{fig:Motivating Example Code}a. The efficient method simply splits based on ``@'', and a comma is added at the second position of each string.  Compared to the efficient method, the inefficient solution not only involves more string operations, such as count and slice, but also utilizes complex regular expression matching, resulting in a runtime difference of an order of magnitude between the two methods. In this scenario, \textbf{making local modifications} doesn't effectively enhance the efficiency of the code. Theoretically, since regular expressions are integral to the inefficient code's structure, traditional performance optimization methods that employ local improvements seem incapable of entirely eliminating dependency on regular expressions to achieve a more optimized solution.

\textbf{Merely} providing \textbf{inefficient code} cannot satisfy the needs of \textbf{efficiency optimization}. Without understanding the NL description of the code's functionality, directly viewing the inefficient code makes it hard to intuitively grasp optimization strategies. Figure \ref{fig:Motivating Example Code}b shows the performance of ChatGPT-4 when only given the inefficient code and asked to optimize its efficiency. Although the code appears quite concise, the actual code generated by ChatGPT-4 is not ideal, presenting two major issues: 1) The inability to move away from the algorithmic framework using regular expressions, and 2) Failure to maintain code functionality. Furthermore, after multiple sampling tests, ChatGPT-4 merely keeps altering the content of the regular expressions without substantive improvements. A significant problem is that these codes employing regular expressions all \textbf{lose} some of their \textbf{functionality}. Notably, when facing the first I/O unit test input ``a@aa@a'' in Figure \ref{fig:Motivating Example}, all the codes using regular expressions fail to output the correct answer ``a@a,a@a''. Here are some regular expression samples generated by ChatGPT-4:
\begin{itemize}
\item \verb|(\b\w+@\w+\.\w+\b)|
\item \verb|[a-zA-Z0-9._-]+@[a-zA-Z0-9._-]+\.[a-zA-Z0-9._-]+|
\end{itemize}

The \textbf{NL description} of the code's functionality \textbf{significantly aids} the optimization task. As illustrated in Figure \ref{fig:Motivating Example Code}c, when ChatGPT-4 is provided with the inefficient code, problem description, algorithm tags, input formats, output formats, and I/O sample tests, its performance is markedly improved. Evidently, with the NL description of the code's functionality, ChatGPT-4 successfully avoids the regular expression-based algorithmic framework, leaning towards a more efficient algorithmic strategy. However, concerning code functionality, although significant progress has been made and it passes most I/O tests, there are still some extreme or specific scenarios in I/O tests, such as ``w@@e'', that are not passed smoothly.

When \textbf{individually} notified about a failed \textbf{I/O test}, further enhancements can be made to the code's functionality. Based on the code generated in Figure \ref{fig:Motivating Example Code}c, we specifically informed ChatGPT-4 that the code failed to process the ``w@@e'' input test. Figure \ref{fig:Motivating Example Code}d displays ChatGPT-4's adjusted results, where an if-statement ``@@ in s'' was added to correct the code. The modified code successfully passed all the I/O unit tests. This discovery suggests that providing specific I/O test feedback to the model individually is highly effective.

\subsection{Key Ideas}\label{sec:3.2}

In the aforementioned example, we identified two primary challenges currently faced in performance optimization: 1) enhancing the flexibility of optimization strategies; and 2) ensuring the maintenance of code functionality.

\textbf{Addressing Challenge 1:} To enhance the adaptability of the generated code, we adopted a code-generation method and introduced the EPC task. By additionally providing an NL description of the code's functionality, we aim to assist the model in generating more efficient code.

\textbf{Addressing Challenge 2:} To ensure the integrity of the code's functionality, we adopted the coding mechanism of an expert encoder group. This strategy achieves a more in-depth and comprehensive information acquisition by extracting features from different input parts separately. Furthermore, given that individual I/O test feedback can notably enhance the accuracy of code generation, an expert encoder, specifically designed for I/O test feature extraction and being part of the expert encoder group, is expected to play a pivotal role in this aspect.

\begin{figure}
    \centering
    \includegraphics[width=0.8\linewidth]{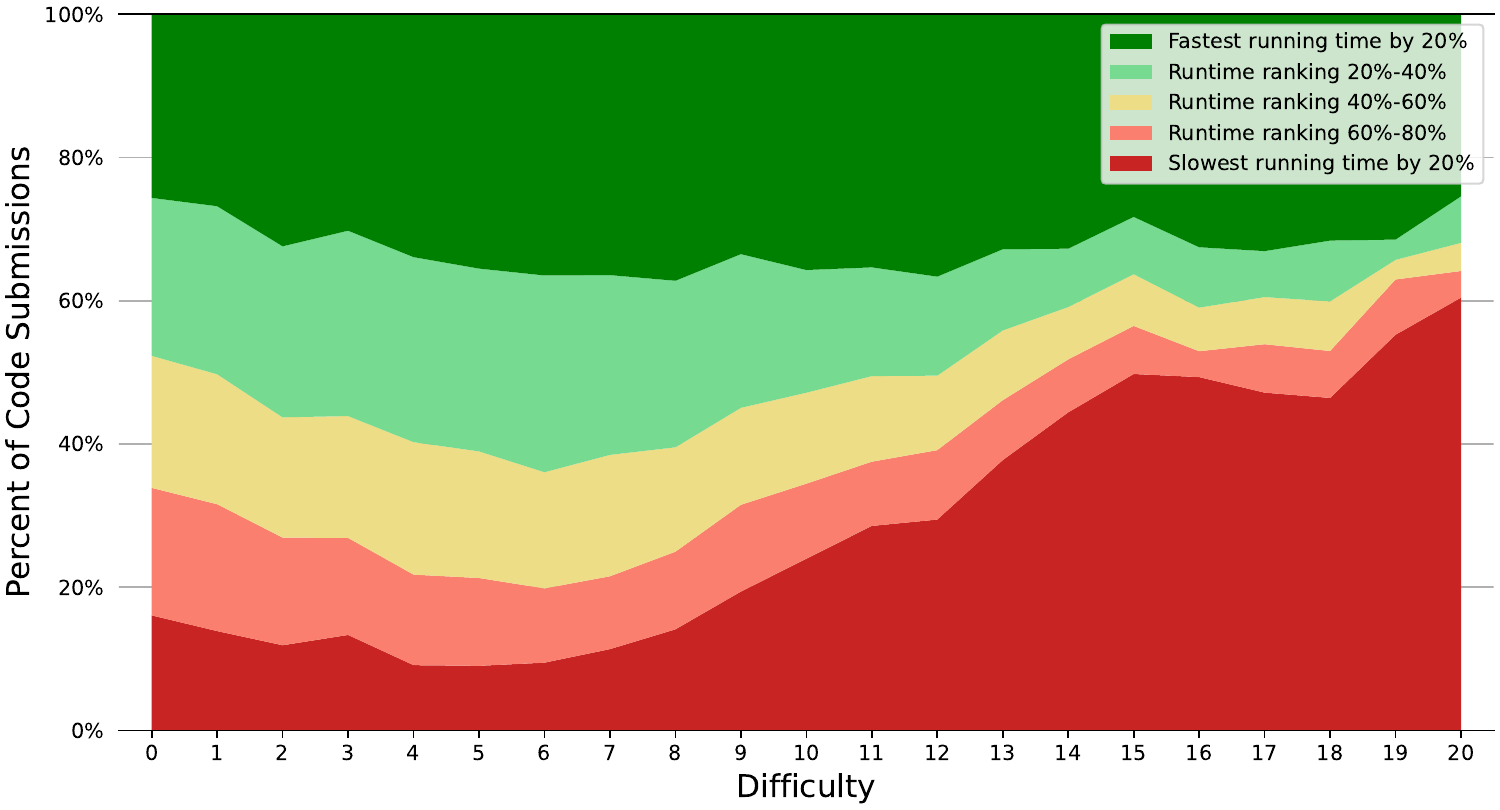}
    \caption{Shows the proportion of codes within the five running time intervals as problem difficulty increases. Figure \mbox{\ref{fig:Time_slice}} displays the method of dividing the five time intervals.}
    \label{fig:Human Code Statistical Chart}
\end{figure}

\begin{figure}
    \centering
    \includegraphics[width=0.6\linewidth]{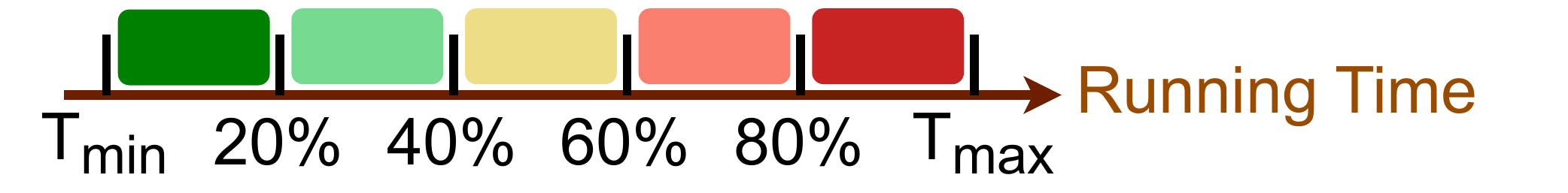}
    \caption{Illustrates the method of segmenting the running time interval for solution codes of the same programming problem, with $T_{min}$ and $T_{max}$ representing the minimum and maximum running times, respectively.}
    \label{fig:Time_slice}
\end{figure}

\section{Efficient Programming Competition Task}\label{sec:4}

\textbf{Traditional programming competitions} often do \textbf{not place sufficient emphasis} on the \textbf{efficiency} of code execution. Programming competitions are a mainstream method to test programming skills, allowing programmers to demonstrate their analytical and problem-solving abilities. While these competitions serve as a predominant approach to evaluate a programmer's analytical and problem-solving capabilities, the primary focus is on how contestants translate an NL problem description into code that meets certain specifications. Even though these competitions do set certain running time constraints for the generated code, the actual time standards are relatively lenient. Based on our analysis of the codes from the CodeForce competition website, we observed that there's a twofold difference in running time between the top 10\% of codes and the median codes; the disparity in running time between the median codes and the bottom 10\% reached five times. This finding suggests that the majority of contestants' codes barely meet the minimum running time requirements, without fully pursuing code efficiency.

Studying human programming habits is crucial for a deep understanding of the mechanisms behind LLMs. Some scholars suggest that the learning process of neural networks may essentially be a form of data compression. In their recent research, Yu et al. \mbox{\cite{yu2023white}} proposed that current AI systems, including GPT-4, primarily engage in compression. They validated this through a newly introduced deep network architecture, CRATE, using mathematical methods. Currently, the learning capabilities of LLMs are mainly limited to extracting and compressing 'code patterns' from large-scale training datasets. These patterns do not fully reach the level of understanding that human programmers achieve when writing efficient code. Given that LLMs are trained based on code written by humans, these models reflect certain characteristics and limitations of human programming. From this perspective, studying human programming habits can help optimize existing model training methodologies, thereby advancing the field of software engineering.

We extracted a dataset comprising 800,000 code entries that comply with the runtime constraints from the CodeForce competition platform. We visualized this data using a stacked area chart, as illustrated in Figure \mbox{\ref{fig:Human Code Statistical Chart}}. 
Specifically, on the CodeForces website, for a given programming problem, we uniformly sampled up to 500 Python codes at equal intervals based on their running times sorted in ascending order. To mitigate the impact of outliers, we removed the top 1\% of extreme running time data. As shown in Figure \mbox{\ref{fig:Time_slice}}, for a given programming problem, we defined a time interval between the minimum running time $T_{min}$ and the maximum running time $T_{max}$ of all solution codes, and divided it into five equal parts by setting six time points. For example, codes with running times within the first part of the interval are classified as “Fastest running time by 20\%”. Specifically, we defined six time points  $T_{i}$ (where i=[0,1,2,3,4,5]) corresponding to [0\%, 20\%, 40\%, 60\%, 80\%, 100\%], as per the following formula:
\begin{equation}
T_{i}= T_{min}  + i \times \frac{(T_{max} -T_{min})}{5}
\end{equation}
where $T_{0} = T_{min}$ and $T_{5} = T_{max}$.

In Figure \mbox{\ref{fig:Human Code Statistical Chart}}, the statistical value of difficulty $d$ represents the average proportion of codes within each interval across all problems of that difficulty level. For the $n$th problem of difficulty $d$, we calculated the total number of codes $C_{total } ^{d-n}$ and the number of codes within each interval $[T_i,T_{i+1}]$ as $C_{ [T_i,T_{i+1}] } ^{d-n}$. The proportion $R_{ [T_i,T_{i+1}] } ^{d}$ for each 20\% interval (e.g., 0\%-20\%) within difficulty $d$ can be calculated using the following formula:
\begin{equation}
R_{ [T_i,T_{i+1}] } ^{d} = \frac{1} {n}  \sum_{n=1} ^{N}  \frac{C_{ [T_i,T_{i+1}] } ^{d-n}}{C_{total } ^{d-n}}
\end{equation}
where $N$ is the number of problems within difficulty $d$.

As shown in Figure \mbox{\ref{fig:Human Code Statistical Chart}}, only approximately 30\% of the codes are classified under “Fastest running time by 20\%,” suggesting that these codes might be challenging to optimize further. However, the remaining codes, while meeting the runtime requirements, still possess substantial potential for efficiency improvements. Moreover, as the complexity of the problems increases, the proportion of the least efficient codes also grows. This trend indicates that most contestants' codes merely satisfy the minimal runtime criteria without fully pursuing code efficiency, a phenomenon that becomes more pronounced with more complex problems.

\begin{figure}
    \centering
    \includegraphics[width=0.95\linewidth]{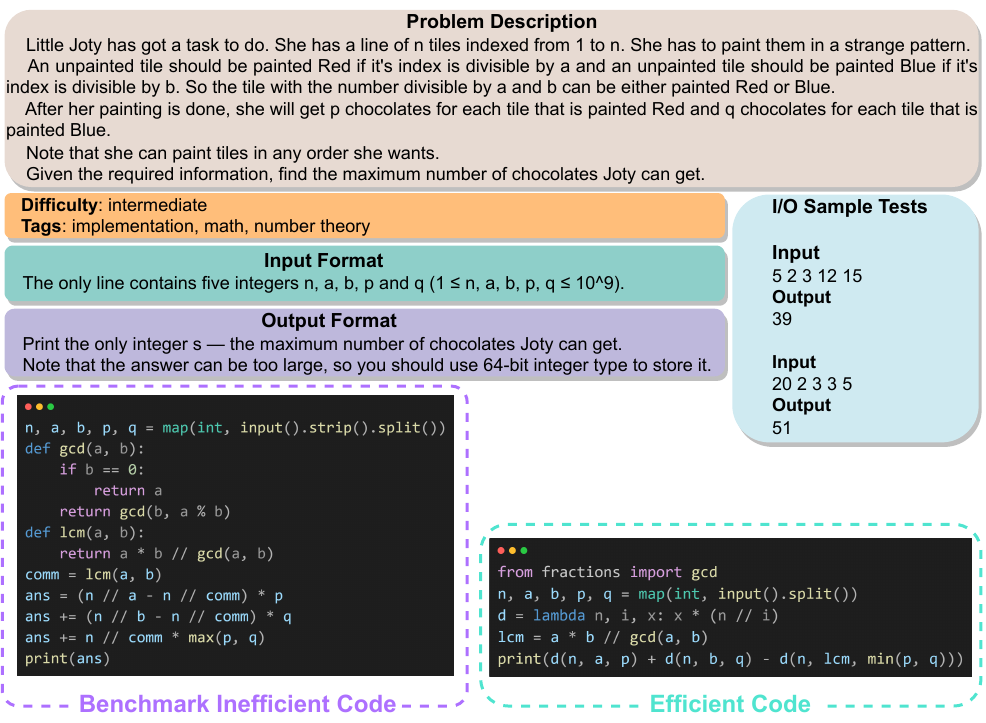}
    \caption{A specific example of the EPC task. This includes the problem's NL description (top), the benchmark inefficient code (bottom left), and the example efficient code (bottom right).}
    \label{fig:EPC_Task}
\end{figure}

To address this issue, we introduced the \textbf{EPC task}, a subclass of the code-generation task. Figure \ref{fig:EPC_Task} presents a specific example of the EPC task. In this task, contestants must understand the NL description of the problem and reference a functionally correct yet inefficient code, with the objective of generating a more efficient version. The bottom right of Figure \ref{fig:EPC_Task} showcases an example of efficient code, optimized by using loops instead of recursion, built-in functions in place of custom functions, and reducing variable naming.

The \textbf{EPC} task \textbf{differs} significantly in its assessment \textbf{objectives} compared to \textbf{traditional programming competitions.} The former primarily focuses on evaluating contestants' skills in translating an NL description into computer code; the latter emphasizes the efficiency of code execution. To steer contestants toward enhancing code efficiency, we provide a benchmark code in the EPC task, which is functionally complete but has a lower execution efficiency. Thus, contestants can avoid spending excessive energy on translating the problem and instead focus more on enhancing code efficiency.

The EPC task can be divided into three main phases:
\begin{itemize}
\item \textbf{Problem Analysis Phase.} Contestants need to carefully read and comprehend the NL description of the problem, which covers algorithm labels, detailed problem description, input format, output format, and associated I/O test examples.
\item \textbf{Algorithm Construction Phase.} In this phase, contestants are required to design and implement an appropriate algorithm to address the given problem. This mandates contestants to possess an in-depth understanding of algorithms and data structures. It's worth noting that the provided benchmark inefficient code serves this phase, aiding contestants to bypass initial coding and directly proceed to code optimization.
\item \textbf{Code Optimization Phase.} After obtaining all essential information, contestants will invest a significant amount of effort to craft the most efficient code possible.
\end{itemize}

\section{Approach}\label{sec:5}

\subsection{Model Architecture}\label{sec:5.1}

\begin{figure}
    \centering
    \includegraphics[width=0.9\linewidth]{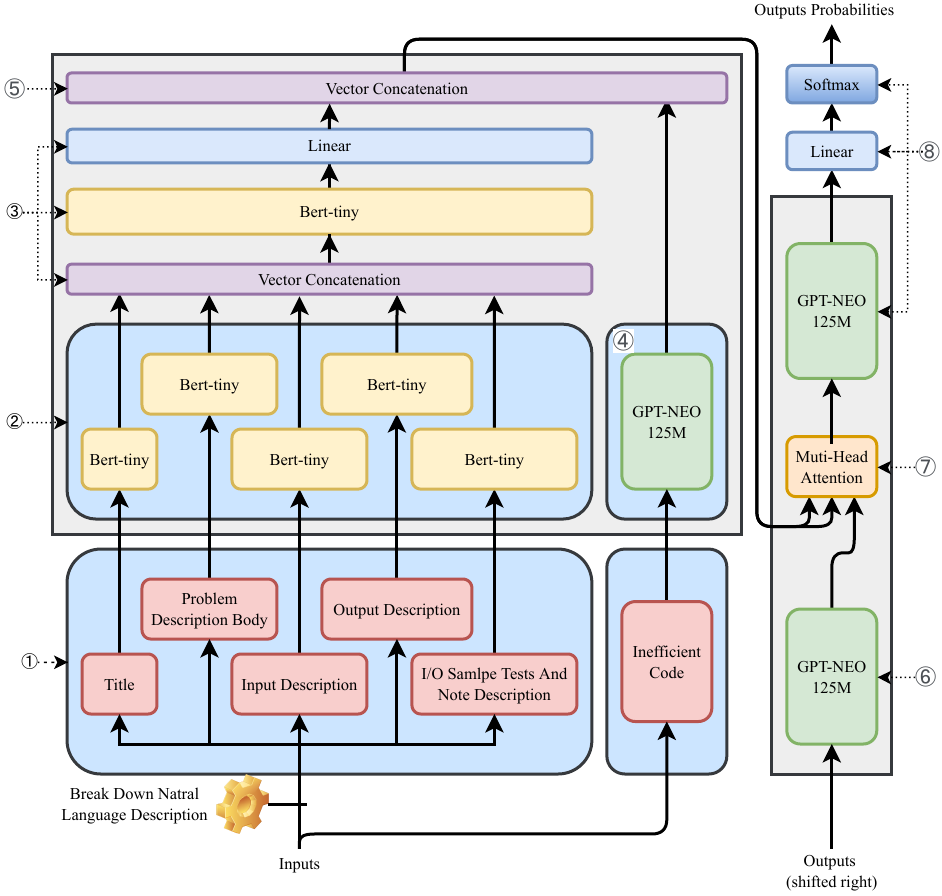}
    \caption{Overview of E-code. Each step has been marked with a numerical symbol (e.g., \textcircled{1}\textcircled{2}\textcircled{3}).}
    \label{fig.E-code}
\end{figure}

We designed and implemented \textbf{E-code}, a model constructed by \textbf{integrating multiple pre-trained models}. An overview of the E-code's structure is presented in Figure \ref{fig.E-code}. Specifically, we incorporated two pre-trained models, Bert-tiny \cite{turc2019well,bhargava2021generalization} and GPT-Neo \cite{black2021gpt}. To ensure their effective integration, extensive structural adjustments were made to the original models, and a new connection mechanism was introduced. By harnessing the power of pre-trained models, we can address the scarcity of competitive programming data and substantially reduce the computational burden during model training.

E-code adopts an \textbf{encoder-decoder architecture}. Existing literature has demonstrated that such an architecture is particularly suited for code generation tasks, a finding further validated by our experimental data. It is noteworthy that the architecture of E-code employs an asymmetric design. The expert encoder group consists of five Bert-tiny models, each configured with a maximum token length of 512. The Expert group integration layer is set to handle 2048 tokens. In contrast, the decoder utilizes only 768 tokens. In total, E-code has 401M parameters.

E-code generates efficient code in an \textbf{autoregressive} manner. The probability of generating efficient code equals the product of the probabilities of each token in the generated code. Specifically, this probability can be interpreted as the likelihood of each token in the generated code appearing in a particular sequence. This process is mathematically represented as:
\begin{equation}
p(EC) = {\textstyle \prod_{t=1}^{P}} p(t_{i} | \text{NL input, IC input, } t_{1},\dots ,t_{i-1})
\end{equation}
where $t_{i}$ is the $i$th token in the generated code, NL input is the input NL description, IC input is the inefficient input code, EC input is the generated efficient code.

In this way, our task is to train a model to compute $p(t_{i} | \text{NL input, IC input, } p_{i})$. That is, given the NL description, inefficient code, and the currently generated portion of the efficient code, the model will compute the probability of the next efficient code token to be generated.

The main processes of the E-code model, which we have labeled with numerical sequences (e.g., \textcircled{1}\textcircled{2}\textcircled{3}) in \ref{fig.E-code}, are as follows:
\begin{itemize}
\item \textbf{Step 1: Partitioning NL descriptions.} We divide the NL description into five parts: algorithm tags, problem descriptions, I/O format descriptions, and I/O test samples. Then, we input them into the expert encoder group. We describe our expert encoder group in detail in \ref{sec:5.2}. Its normalized expression is:
\begin{equation}
nl_t,nl_q,nl_i,nl_o,nl_s=f^{(Split)}(\text{NL input})
\end{equation}
where $f^{(Split)}$ is the segmentation function, $nl_t,nl_q,nl_i,nl_o,nl_s$ is the NL description of the segmentation into five parts.

\item \textbf{Step 2: Expert group processing.} The expert encoder group consists of five Bert-tiny models \cite{turc2019well,bhargava2021generalization}, each Bert-tiny model dealing with a separate section. Since the Bert-tiny models are small, they fit into our expert encoder group. The feature information set $e_X$ extracted by the expert group is computed by
\begin{equation}
e_{X} = f_{X}^{(Bert)}(nl_{X}),\text{ } X\in  \{t,q,i,o,s\}
\end{equation}
where $f_{X}^{(Bert)}$ is the set of expert group functions.

\item \textbf{Step 3: Expert group integration layer integration.} The output of the expert encoder group is concatenated and integrated using Bert-tiny models \cite{turc2019well,bhargava2021generalization} so that split NL descriptions can focus on each other and extract deep-level features. These features are then fed into a layer of \textbf{M}ulti-\textbf{L}ayer \textbf{P}erceptron (MLP) layers for integration. Note that our MLP layer deliberately removes the ReLU activation function. We describe this in detail in \ref{sec:5.3}. The NL features $Enc^{(nl)}$ of the encoder is computed by
\begin{equation}
Enc^{(nl)} = W^{(Enl)} [f_{(Int)}^{(Bert)} concat(e_t , e_q , e_i , e_o , e_s)]
\end{equation}
where $concat$ is the operator of tensor concatenation, $f_{(Int)}^{(Bert)}$ is the expert group integration layer function, $W^{(Enl)}$ is the MLP enlarge layer weights.

\item \textbf{Step 4: Extracting information about inefficient code features.} The GPT-Neo model \cite{black2021gpt} is used to extract feature information from the inefficient code. Because the GPT-Neo model has been pre-trained for the code, we can reduce the computational overhead of fine-tuning again. Inefficient code features information $Enc^{(ic)}$ computed by
\begin{equation}
Enc^{(ic)} =f_{ic}^{(GPT-Neo)}(\text{IC input})
\end{equation}
where $f_{ic}^{(GPT-Neo)}$ is the GPT-Neo model for extracting information about inefficient code features.

\item \textbf{Step 5: Integrated encoder output.} Concatenate the features obtained by the expert group with the components extracted from the inefficient code as the output of the encoder.  The encoder output $Enc$ is computed by
\begin{equation}
Enc= concat(Enc^{(nl)},Enc^{(ic)})
\end{equation}

\item \textbf{Step 6: Extracting information about generated efficient code features.} Feature information is extracted from the generated efficient code using the GPT-Neo model.  The extracted efficient code feature information $Dec^{(ec)}$ is computed by
\begin{equation}
Dec^{(ec)} =f_{ec}^{(GPT-Neo)}(\text{EC input})
\end{equation}
where $f_{ec}^{(GPT-Neo)}$ is the GPT-Neo model for extracting information about efficient code features.

\item \textbf{Step 7: Multi-headed attention mechanism.} Integrate the feature information of the encoder output and the generated partially efficient code using the multi-head attention layer and feed it to the final output module. Since the model decoder is unique, the attention mechanism layers in the model are independent. All our attention layers use 48 headers to maximize the fusion of feature information. We explain in detail the reasons for using 48 headers in Section \ref{sec:5.3}. The multi-head attention layer output $Dec^{(multi-h)}$ is computed by
\begin{equation}
Dec^{(multi-h)} = concat(head_1,\dots ,head_H ) W_h
\end{equation}
where $H$=48 denotes the number of heads, $W_h$ is the weight. An attention layer is applied in each head $head_t$, computed by
\begin{equation}
head_t=softmax(QK^\top/{\sqrt{d_k}} )V
\end{equation}
where $d_k = d/H$ denotes the length of each features vector. $Q$, $K$ and $V$ are computed by
\begin{equation}
[Q,K,V]=[Dec^{(ec)},Enc,Enc]^\top  [W_Q ,W_k ,W_V]
\end{equation}
where $W_Q ,W_k ,W_V$ are model parameters.

\item \textbf{Step 8: Predicted output.} The GPT-Neo model \cite{black2021gpt} is again used as the final output module to predict the next token based on all the feature information. The probability of predicting the next token $p(t_i)$ is computed by
\begin{equation}
p(t_i)=softmax(f_{output}^{(GPT-Neo)}(Dec^{(multi-h)}))
\end{equation}
where $f_{output}^{(GPT-Neo)}$ is the GPT-Neo model with MLP layers.
\end{itemize}

E-code possesses \textbf{robust capability} in generating efficient code. Our E-code uses the expert encoder group to process distinct parts of the inefficient code and problem NL description separately. Efficient and robust feature extraction is a prerequisite for producing efficient code; utilizing guidance from the inefficient code ensures the development of efficient output. By establishing a mapping relationship between inefficient and efficient code, and through gradient propagation using high-quality efficient code as the ground truth, our E-code can ultimately generate satisfactory efficient code.

\subsection{Expert Encoder Group}\label{sec:5.2}

\textbf{The expert encoder group} was specifically designed to handle tasks with \textbf{multiple types} of \textbf{inputs}. In practical software performance optimization work, it is often necessary to integrate and extract various information features. Similarly, in the EPC task, multiple types of input data exist. The expert encoder group can extract features from various input information through different expert encoders. More importantly, this structure can flexibly adapt to different types of information changes by adding or removing expert encoders.

In this paper, the expert encoder group comprises \textbf{six expert encoders}, designed to extract \textbf{different types} of \textbf{information features} for the EPC task. The group includes five Bert-tiny models and one GPT-Neo model \mbox{\cite{black2021gpt}}, each employed to extract features from different types of outputs. Specifically, the fine-tuned GPT-Neo model is utilized to capture features of inefficient code, while the five Bert-tiny models \mbox{\cite{turc2019well,bhargava2021generalization}} are each used to extract features from algorithm tags, problem descriptions, input formats, output formats, and I/O test samples. Ultimately, all extracted features are integrated through the Expert Group Integration Layer.

The expert encoder group exhibits robust generalization capabilities, key to handling diverse input types. This is achieved by employing distinct expert encoders to process various types of model inputs, fundamentally leveraging specifically trained models to manage different input types. This approach allows tasks with varied types of inputs at the input end to be feasibly addressed by the expert encoder group.

The expert encoder groups have achieved commendable results in feature extraction. Our design decisions were primarily to take full advantage of this expert group. We summarized three major advantages of the expert encoder group:

\begin{center}
\begin{tcolorbox}[colback=gray!5,%gray background % 背景 颜色
                  colframe=black,% black frame colour % 字体颜色
                  width=16cm,% Use 8cm total width,%文本框宽度
                  arc=2mm, auto outer arc, % 文本框弧度
                  boxrule=0.5pt, %
                  toprule=1.2pt, rightrule=1.2pt, bottomrule=1.2pt,leftrule=1.2pt, % 边框粗细
                  ]

\textbf{High Specialization.} By adopting expert encoder groups, each encoder demonstrates stronger specificity when processing tasks. This effectively reduces the negative task transfer \mbox{\cite{aghajanyan2021muppet, asai2022attempt, zhang2022survey,levine2022standing}} and catastrophic forgetting \mbox{\cite{mccloskey1989catastrophic, chakrabarty2022fine}} that may occur in multi-task learning. In the traditional ``NL+Code'' input mode, the NL part is typically shorter and mostly acts as a comment to supplement the code. This feature is particularly evident in basic programming questions. However, in EPC tasks, the NL portion is usually longer than the code and includes various types. This makes the simplistic approach of treating the NL as merely a code comment inappropriate. When traditional LMs handle EPC tasks, the overall input method causes the elongated NL section and the code section to negatively impact each other, thereby affecting the model's overall performance. Our proposed expert encoder groups can process different input types separately, converting the unfamiliar ``long NL+Code'' input mode for LMs into the more proficiently handled singular ``NL'' or ``Code'' modes. This allows the expert encoder groups to enhance the efficiency of E-code comprehensively, rather than merely achieving improvement in a specific metric.

\end{tcolorbox}
\end{center}

\begin{center}
\begin{tcolorbox}[colback=gray!5,%gray background % 背景 颜色
                  colframe=black,% black frame colour % 字体颜色
                  width=16cm,% Use 8cm total width,%文本框宽度
                  arc=2mm, auto outer arc, % 文本框弧度
                  boxrule=0.5pt, %
                  toprule=1.2pt, rightrule=1.2pt, bottomrule=1.2pt,leftrule=1.2pt, % 边框粗细
                  ]
                  
\textbf{Robust Generalization Capability.} The expert encoder group has high scalability and can flexibly add or remove expert encoders to process various types of information. In performance optimization tasks, the provided information is highly diverse, integrated, and complex. In reality, different users might provide very different information for the same task since there are multiple ways to describe a task. For instance, performance optimization may require extracting features from NL and inefficient code, where the NL can be further broken down into NL descriptions, programming constraints, code comments, etc. Furthermore, task descriptions can also be expressed based on different granularities, such as background knowledge, functionality, and algorithmic processes.

\end{tcolorbox}
\end{center}

\begin{center}
\begin{tcolorbox}[colback=gray!5,%gray background % 背景 颜色
                  colframe=black,% black frame colour % 字体颜色
                  width=16cm,% Use 8cm total width,%文本框宽度
                  arc=2mm, auto outer arc, % 文本框弧度
                  boxrule=0.5pt, %
                  toprule=1.2pt, rightrule=1.2pt, bottomrule=1.2pt,leftrule=1.2pt, % 边框粗细
                  ]
                
\textbf{Low Computational Cost.} Dividing the original encoder into five parts can effectively reduce computational overhead. This is because the computational complexity of the encoder grows exponentially with its length. The expert encoder group is built upon the Transformer model \cite{vaswani2017attention}, where most of the computational resources of the Transformer model are consumed in the Self-Attention part. The time complexity of self-attention is $O(n^2 \cdot d)$, where $n$ denotes the sequence length, and $d$ represents the embedding dimension. This complexity stems from three principal phases: similarity computation, Softmax, and weighted averaging. In theory, if we divide a complete input sequence into five parts of equal length, the computational overhead will be reduced to one-fifth of the original cost, i.e., $5 \times O(1^2 \cdot d) = 1/5 \times O(5^2 \cdot d)$. In contrast, our expert group processes the complete input in five independent parts, effectively reducing computational costs. Moreover, the operation of the Expert Group Integration Layer in Step 3 is identical to that of the standard encoder framework, theoretically resulting in no difference in runtime between the two.

\end{tcolorbox}
\end{center}

\subsection{Integrating Pre-trained Language Models}\label{sec:5.3}

E-code employs \textbf{multiple strategies} to enhance model performance. Earlier research has demonstrated the effectiveness of combining individually fine-tuned language models for distributed multi-task fine-tuning \cite{li2022branch,don2023cold}. Based on this, E-code is composed of several integrated pre-trained LMs. Analogous to the manner in which each expert encoder in the expert encoder group handles a unique task, each pre-trained LM in E-code is also assigned a distinct task. Several strategies are introduced when merging these distributedly trained language models.

\begin{itemize}
\item \textbf{Use of MLP to Alter Tensor Dimensions.} When integrating two models with different tensor sizes, one may encounter dimension-matching issues. To address this challenge, we employed a MLP to adjust the tensor dimensions.
\item \textbf{Removal of the ReLU Activation Function from the MLP Layer.} Although MLP is typically used in conjunction with the ReLU activation function, in this work, we opted to remove the ReLU activation function. The rationale behind this decision is that the primary role of the MLP here is to adjust the tensor's dimension and not to introduce non-linearity. While the ReLU activation function can introduce non-linearity to the MLP, enabling it to better approximate non-linear functions, in this context, non-linearity could lead to feature loss, potentially affecting model performance.
\item \textbf{Use of Attention Mechanism.} Unlike direct connections, the attention mechanism between the encoder and the decoder assists in better preserving the feature structure information. Additionally, it can more effectively integrate features extracted from different models.
\item \textbf{Employing a 48-head Attention Mechanism.} Given that the standard Transformer decoder typically contains six decoding sections, and each section's attention mechanism comprises eight heads, our model, aiming for optimal integration of encoder and decoder information, chooses to use an attention mechanism with 48 heads for improved results.
\end{itemize}

\subsection{ExecTimePredictor}\label{sec:5.4}

To effectively \textbf{predict} the \textbf{running time} of code, we introduced the \textbf{ExecTimePredictor}. In the generation of efficient code, the running time of the code serves as a crucial metric for evaluating its efficiency. A short running time indicates efficient code, while a long running time implies inefficiency. However, a significant issue arises if the generated code cannot compile, making it challenging to ascertain its efficiency. Given the current limitations of LMs in ensuring the functional correctness of generated code, and the fact that code which might be fixed with minor token adjustments still holds potential value, this paper seeks solutions to reasonably assess the efficiency of such code. To address this issue, we introduce ExecTimePredictor, a tool capable of predicting the execution time of uncompileable code.

We employed DeepSeek-Coder \mbox{\cite{guo2024deepseek}}, a cutting-edge code LLM, as the foundational model for ExecTimePredictor. DeepSeek-Coder surpasses existing code LLMs such as Code Llama and Star Coder. We selected version DeepSeek-Coder-v1.5 7B, which is based on DeepSeek-LLM-7B and additionally pretrained on an extra 2 trillion tokens. To ensure ExecTimePredictor's effectiveness on both executable and non-executable code, we created a dataset specifically for fine-tuning ExecTimePredictor. We crawled and preprocessed the necessary code data, as detailed in the following steps:

\begin{itemize}
\item We extracted a large volume of code data from the CodeForces website, which underwent preprocessing such as removing code that could not generate an AST, failed I/O tests, or was excessively long.
\item Given that the runtime data provided by CodeForces was not sufficiently precise, we recalculated the execution times. By wrapping the code in functions and recording the time taken to call these functions, we significantly enhanced the precision of our execution time measurements, converting the units from milliseconds to microseconds.
\item We divided the test set according to the submission times on the website to prevent data leakage.
\end{itemize}

Ultimately, we compiled the CodeExecTimeDB, which contains a total of 147,677 entries, 16,662 of which are designated as the test set. Depending on different needs, we created four distinct dataset variants, as follows:

\begin{itemize}
\item \textbf{CodeExecTimeDB-Ori:} The original version, without any modifications.
\item \textbf{CodeExecTimeDB-Uni:} Builds on CodeExecTimeDB-Ori by standardizing variable and function names (e.g., renaming variables to var1). Since differences in variable and function names do not affect execution time, this helps to reduce noise interference. Additionally, as the GEC dataset is insensitive to variable and function names, the fine-tuned model produces code with uniform naming conventions. Therefore, ExecTimePredictor is trained to handle such code.
\item \textbf{CodeExecTimeDB-Loop\&Rec:} Based on CodeExecTimeDB-Uni, retains only loop and recursion statements. These elements often account for a significant portion of the code’s execution time and remain relevant even if the overall code is non-executable. This variant aims to teach the model to infer the overall code execution time solely from loops and recursion.
\item \textbf{CodeExecTimeDB-RandDel:} Builds on CodeExecTimeDB-Uni by randomly deleting 20\% of code tokens. This random deletion mimics non-executable code more closely, further ensuring ExecTimePredictor's effectiveness on such code.
\end{itemize}

% Theept (see Section \mbox{\ref{7.2}}).

% To \textbf{fine-tuning} the ExecTimePredictor, we constructed a dataset named \textbf{CodeExecTimeDB}. This dataset comprises approximately 1.24 million data entries, of which a random subset is designated as the test set at a ratio of 50:1. CodeExecTimeDB provides both unified and non-unified versions of code variables and function names, and ExecTimePredictor was separately fine-tuned on these two versions. On one hand, given that GEC is a dataset insensitive to variable and function names, the fine-tuned model generates code with uniform variable and function naming. Therefore, the ExecTimePredictor needs to handle such code. On the other hand, differences in variable and function names do not impact the running time of the code. Thus, an ExecTimePredictor insensitive to such differences would yield more accurate predictions.

\subsection{Efficiency Gain Ratio}\label{sec:5.5}

To quantitatively \textbf{assess} the effects of \textbf{code optimization}, we introduced a novel metric, the \textbf{E}fficiency \textbf{G}ain \textbf{R}atio (EGR). EGR calculates the ratio of optimized time, mathematically defined as:
\begin{equation}
\text{EGR} = (T_{\text{baseline}} - T_{\text{enhanced}})/{T_{\text{baseline}}}
\end{equation}
where $T_{\text{baseline}}$ represents the running time of the inefficient code, and $T_{\text{enhanced}}$ denotes the running time of the efficient code. As a standardized measure of performance improvement, EGR facilitates fairer comparisons across different code segments and computational tasks. A higher EGR value suggests a larger efficiency boost due to optimization, whereas values close to zero indicate minimal enhancement. Through the adoption of this metric, our goal is to offer a standardized method to evaluate the impact of code efficiency improvements in real-world scenarios.

\section{Experimental Setup}\label{sec:6}

\subsection{Reserach Questions}\label{sec:6.1}

In this research, we present the experimental design and its outcomes. Specifically, we aim to address the following research questions:

\textbf{RQ1:} How does E-code perform in the EPC task compared to baseline models?

\textbf{RQ2:} How does the performance of E-code differ from that of baseline models in the EPC task for problem NL descriptions of varying difficulty?

\textbf{RQ3:} How much does an increase in code submission frequency influence the improvement of E-code's performance?

\textbf{RQ4:} What are the common failure modes of E-code and ChatGPT in the EPC tasks?

% \textbf{RQ5:} Can E-code provide extensive performance optimization?

\textbf{RQ5:} Are all components of E-code (such as expert encoder groups, unified code variables, function naming, and model concatenation usage) necessary? What is the impact on overall performance when these components are individually ablated in ablation studies?

\subsection{Comparison Baselines}\label{sec:6.2}

To comprehensively evaluate the performance of E-code, we selected \textbf{several representative baseline models} for comparison experiments, covering the common model architectures in the EPC task. Based on the differences in model structure, these baseline models can be divided into two major categories: \textbf{Left-to-Right LMs} and \textbf{Encoder-decoder LMs}. Except for CodeT5-base which belongs to Encoder-decoder LMs, the rest of the LMs fall under Left-to-Right LMs.

\begin{itemize}
\item \textbf{GPT-Neo (125M)} \cite{black2021gpt}. While GPT-Neo might have a smaller parameter size, it still exhibits impressive NL processing capabilities.
\item \textbf{CodeT5-base (220M)} \cite{codet5}. CodeT5-base is a pre-trained model based on the T5 architecture, specially optimized for tasks involving code and NL interaction.
\item \textbf{CodeGen (350M)} \cite{zhong2022codegen}. CodeGen-mono is a code-generation model built upon GPT-2. It not only displays strong performance but also supports multiple programming languages.
\item \textbf{PolyCoder (400M)} \cite{xu2022systematic}. PolyCoder is an innovative deep learning model, focusing on encoding and decoding across multiple programming languages. It supports both automated code generation and program comprehension.
\item \textbf{ChatGPT} \cite{chatgpt}. ChatGPT has showcased powerful prowess in code generation. It can not only comprehend and interpret coding requirements but also produce corresponding code snippets based on these requirements, thus enhancing development efficiency. Especially when handling regular programming tasks and problems, its code prediction and generation accuracy are very high. In this context, we used ChatGPT-3.5-turbo.
\end{itemize}

\subsection{Fine-tuning Setup}\label{sec:6.3}

We utilized the \textbf{GEC} dataset \cite{Yue2023GEC} for \textbf{fine-tune} the EPC task. The GEC contains all the necessary information required for the EPC task, with a total of 31,577 data entries, out of which 3,085 were allocated to the test set. Both E-code and all baseline models were fine-tuned on the GEC training set and assessed on its test set. In the fine-tuning phase, the NL description of the problem and the inefficient code are considered as the ``features'' inputs to the model, while the efficient code serves as the ``label'' for gradient updating. Notably, the ``label'' code not only has the shortest running time but also requires less space during execution. This ensures that the model doesn't tend to sacrifice space to save time.

Throughout the fine-tuning process, we used the AdamW loss function \cite{loshchilov2017decoupled}, set a batch size of 32, and underwent 15 fine-tuning epochs. As Austin et al. \cite{austin2021program} found that the sampling method is superior to the beam search method, we adopted top-k=50 \cite{fan2018hierarchical} and nucleus sampling with top-p=0.95 \cite{holtzman2019curious}. Temperature tuning, a regularization technique proposed by Dabre and Fujita, artificially renders the label probability distribution smoother or sharper \cite{chen2021evaluating}. We referred to the temperature settings from the AlphaCode system \cite{li2022competition} and the CodeX model \cite{dabre2020softmax} and used a temperature of 0.25. Aside from this, other model parameter configurations adhered to the default settings by Hugging Face. Additionally, during the fine-tuning process of ExecTimePredictor, we set the training epochs to 10, with other settings remaining unchanged. ExecTimePredictor's prediction error on the CodeExecTimeDB test set was only \textbf{20.99 ms}, which is within acceptable bounds. When using LLMs for fine-tuning and predicting the EPC task, we designed the following prompt to instruct the model on what to generate:
\begin{center}
\begin{tcolorbox}[colback=gray!5,%gray background % 背景 颜色
                  colframe=black,% black frame colour % 字体颜色
                  width=14.5cm,% Use 8cm total width,%文本框宽度
                  arc=2mm, auto outer arc, % 文本框弧度
                  boxrule=0.5pt, %
                  toprule=1.2pt, rightrule=1.2pt, bottomrule=1.2pt,leftrule=1.2pt, % 边框粗细
                  ]

``Natural Language Description:\textbackslash n\{\textit{\textbf{Natural Language Description}}\}\textbackslash nInefficient code:\textbackslash n\{\textit{\textbf{Inefficient Code}}\}\textbackslash nPlease produce an optimized version:''

\end{tcolorbox}
\end{center}

\subsection{Fine-tuning ExecTimePredictor}\label{sec:6.4}

To develop the ExecTimePredictor, we fine-tuned the DeepSeek-Coder-v1.5 7B using the AdamW optimizer with a batch size of 32. The model was fine-tuned over four datasets, each for two epochs. After this process, we obtained the refined ExecTimePredictor. The fine-tuned ExecTimePredictor demonstrated accuracies of 8.50 µs, 8.11 µs, 8.35 µs, and 8.21 µs on the CodeExecTimeDB-Ori, CodeExecTimeDB-Uni, CodeExecTimeDB-Loop\&Rec, and CodeExecTimeDB-RandDel test datasets, respectively.

During the fine-tuning and prediction phases with DeepSeek-Coder-v1.5 7B, we designed the following prompt, inspired by the chat template provided by DeepSeek, to guide the model in predicting the execution time of code:

\begin{center}
\begin{tcolorbox}[colback=gray!5,%gray background % 背景 颜色
                  colframe=black,% black frame colour % 字体颜色
                  width=14.5cm,% Use 8cm total width,%文本框宽度
                  arc=2mm, auto outer arc, % 文本框弧度
                  boxrule=0.5pt, %
                  toprule=1.2pt, rightrule=1.2pt, bottomrule=1.2pt,leftrule=1.2pt, % 边框粗细
                  ]

``\#\#\# Code Solution:\textbackslash n\{\textit{\textbf{Code}}\}\textbackslash n\textbackslash n\#\#\# Predict the running time of the provided code solution:''

% ``Natural Language Description:\textbackslash n\{\textit{\textbf{Natural Language Description}}\}\textbackslash nInefficient code:\textbackslash n\{\textit{\textbf{Inefficient Code}}\}\textbackslash nPlease produce an optimized version:''

\end{tcolorbox}
\end{center}

\subsection{Metrics}\label{sec:6.5}

To gauge the model's capability in the EPC task, we mainly examined it from both \textbf{functionality} and \textbf{efficiency} perspectives. The EPC task aims to produce code that's both efficient and functionally accurate. Among these, efficiency clearly trumps functionality. Even if a code snippet contains certain token errors, as long as its algorithmic logic is efficient, it remains of referential significance to programmers. For a comprehensive assessment of the code's functionality and efficiency, we adopted the following four evaluation metrics:
\begin{itemize}
\item \textbf{CodeBLEU.} Compared to the conventional BLEU score, CodeBLEU is an evaluation criterion optimized specifically for the code generation domain. It's worth noting that we employed a calculation method consistent with IOCCB \cite{Yue2023GEC}. For each data entry, we considered multiple ground truth codes, eventually selecting the one with the highest CodeBLEU score.
\item \textbf{EGR.} The EGR serves as a standardized measurement for performance improvements, allowing for a fairer comparison across different code snippets and programming challenges. To minimize noise, including for ground truth code, all code running time were predicted using ExecTimePredictor.
\item \textbf{Compilation.} The compilation rate reflects the model's grasp of code syntax, making it a critical metric for evaluating code quality.
\item \textbf{I/O.} When the code successfully passes all I/O test cases, we consider it functionally correct.
\end{itemize}

\section{Result and Analysis}\label{7}

\subsection{RQ1: Comparison with Baselines}\label{sec:7.1}

E-code \textbf{outperforms} other state-of-the-art models in terms of \textbf{efficiency}. Table \ref{tab:BaseLine} showcases the performance of E-code and all baseline models on the EPC task when the number of code submissions is one. The results indicate that E-code surpasses other models in code execution efficiency and only slightly falls behind ChatGPT in terms of code functionality. Moreover, its higher CodeBLEU score suggests that rectifying its functionality might only require minor token modifications. This further attests to the superiority of E-code.

\begin{table}[htbp]
\caption{The performance of E-code and all baseline models in EPC tasks when the number of code submissions is one.}
\begin{center}
\begin{tabular}{@{}lcccc@{}}
\hline

\textit{\textbf{Models}}               & \textit{\textbf{CodeBLEU}}      &   \textbf{\textit{EGR}}      &     \textbf{\textit{Compilation}}    &     \textbf{\textit{I/O}}     \\ \midrule

\textbf{GPT-Neo}                           & 4.16  &	 18.45\%  &	1.25\%  &	0              \\ 

\rowcolor{gray!10} \textbf{CodeGen}        & 4.68  &	 3.82\%  &	0.4\%  &	0             \\

\textbf{PolyCoder}                           & 2.67  &	11.85\%  &	0.36\%  &	0                 \\

\rowcolor{gray!10} \textbf{CodeT5}           & 10.69  &	 23.29\%  &	1.69\%  &	0.13\%                        \\

\hdashline 

\textbf{ChatGPT}                              &    10.4  & 22.09\%  &	\textbf{87.16\%}  &	\textbf{37.21\%}                            \\

\hline

\rowcolor{gray!10} \textbf{E-code}          &  \textbf{14.44}  & \textbf{30.74\%}  &	76.94\%  &	0.93\%                \\ 

\hline
\end{tabular}
\label{tab:BaseLine}
\end{center}
\end{table}

\textbf{Simply through fine-tuning}, some models did \textbf{not fare well} on the EPC task. After straightforward fine-tuning with GEC, it seems that some models still fail to learn how to optimize code efficiency. This implies that mere fine-tuning might not be sufficient for models to fully grasp the knowledge related to code efficiency. To achieve this aim, guiding the model might be necessary right from the pre-training phase.

\begin{figure}
    \centering
    \includegraphics[width=0.95\linewidth]{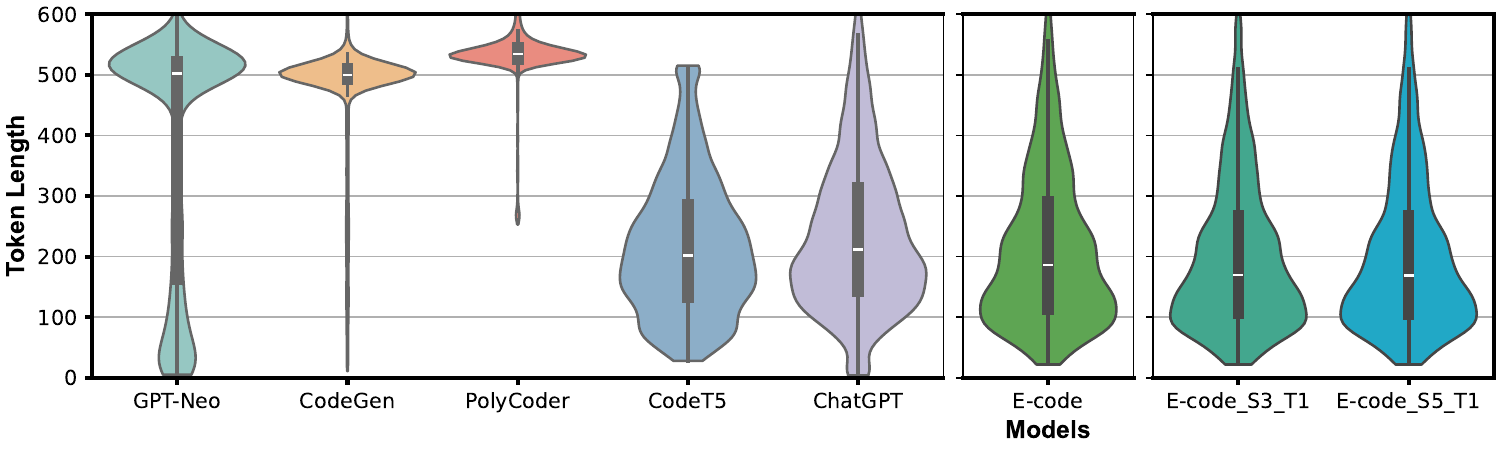}
    \caption{The left half illustrates the token length distribution of code generated by baseline models and E-code. The right half highlights the impact of the filtering mechanism on the token length distribution of code generated by E-code.}
    \label{fig:Code_Token}
\end{figure}

\begin{figure}
    \centering
    \includegraphics[width=0.95\linewidth]{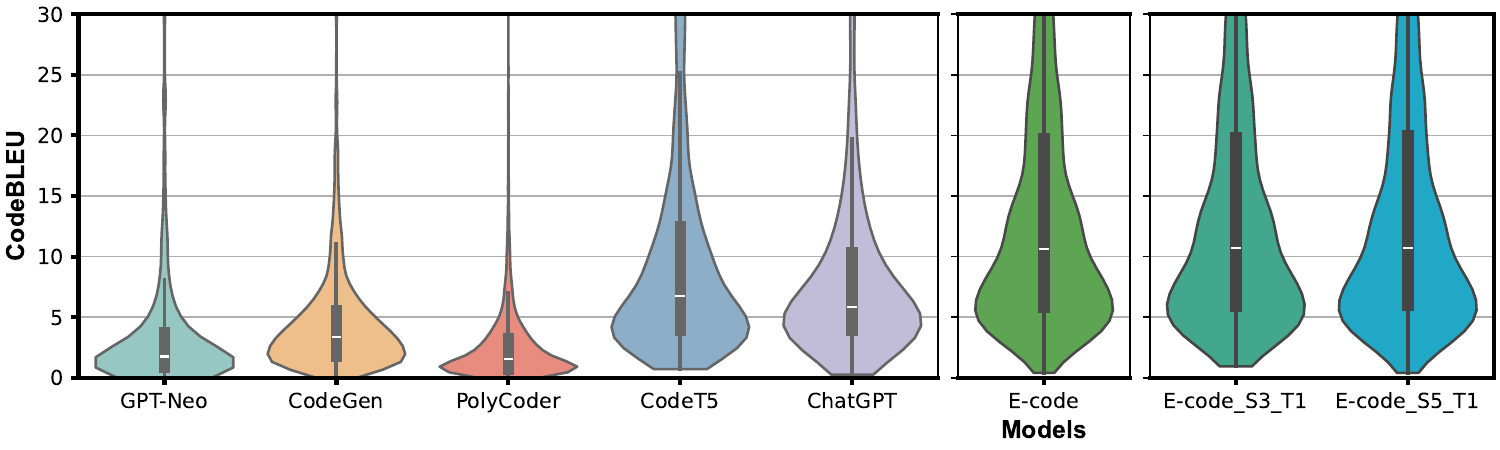}
    \caption{The left half portrays the CodeBLEU distribution of code generated by baseline models and E-code. The right half reveals the influence of the filtering mechanism on the CodeBLEU distribution of code produced by E-code.}
    \label{fig:CodeBLEU}
\end{figure}

Figures \mbox{\ref{fig:Code_Token}} and \mbox{\ref{fig:CodeBLEU}} display the data visualized as violin plots, a type of graph that integrates features of both box plots and density plots to illustrate the shape, density, and probability density function of the data distribution. These plots provide more comprehensive information than either box plots or density plots alone. The central black line within each violin plot represents the data's quartiles (Q1, Q2—the median, and Q3), with the thickest part denoting the median. The shape of the violin indicates the density of data distribution; wider sections suggest a higher concentration of data points, whereas narrower sections indicate fewer data points.

E-code demonstrates more \textbf{consistent} performance. The left half of Figure \ref{fig:Code_Token} displays the token length distribution of the code generated by E-code and the baseline models. The data reveals that decoder models often generate excessively long code, while encoder-decoder models produce code of a more reasonable length. Additionally, in comparison to ChatGPT, the code generated by E-code is shorter. We believe that this shorter code might be the reason why the code generated by E-code underperforms in terms of functionality. Moreover, the appropriate code length closely correlates with the performance across various metrics. The left halves of Figures \ref{fig:CodeBLEU} showcase the CodeBLEU distributions of the code generated by E-code and the baseline models, respectively. The results depict that the CodeBLEU distribution of E-code is more uniform, and it exhibits greater stability.

\subsection{RQ2: Impact of NL Problem Difficulty on Code Functionality and Efficiency}\label{sec:7.2}

The \textbf{difficulty} of the problem's NL description \textbf{inversely correlates} with the \textbf{functionality} and \textbf{efficiency} of the code. Within GEC's evaluation framework, based on the difficulty of the problem's NL description, data is categorized into ``easy'', ``intermediate'' (abbreviated as inter.), and ``hard''. Table \ref{tab:Difficulty} displays the performance of various models across these different NL difficulty classifications. Observing the overarching trend, the difficulty of the problem's NL description generally inversely relates to code functionality and efficiency, with efficiency being more substantially affected than functionality. Our analysis suggests that the difficulty of the problem's NL description primarily gauges the complexity of transitioning from the problem description to the code. However, in efficient programming competitions, participants are predominantly tasked with producing efficient code and are already furnished with a reference code.

\begin{table}[htbp]
\caption{The performance of various models across different NL difficulty classifications.}
\begin{center}
\resizebox{0.9\textwidth}{!}{%
\begin{tabular}{@{}l c c c c c c c c c c c c@{}} %使用“tabularx”环境，并指定宽度为文章宽度
\hline
                    &   \multicolumn{3}{c}{\textbf{CodeBLEU}}
                    &   \multicolumn{3}{c}{\textbf{EGR}}
                    &   \multicolumn{3}{c}{\textbf{Compilation}}  
                    &   \multicolumn{3}{c}{\textbf{I/O}}     \\
                    
\cmidrule(l){2-4}   \cmidrule(l){5-7}     \cmidrule(l){8-10}    \cmidrule(l){11-13}

\multirow{-2}{*}{\textit{\textbf{Method}}} 
&    {\textbf{\textit{Easy}}}    &    {\textbf{\textit{Interm.}}}   &    {\textbf{\textit{Hard}}} 
&    {\textbf{\textit{Easy}}}    &    {\textbf{\textit{Interm.}}}   &    {\textbf{\textit{Hard}}} 
&    {\textbf{\textit{Easy}}}    &    {\textbf{\textit{Interm.}}}   &    {\textbf{\textit{Hard}}} 
&    {\textbf{\textit{Easy}}}    &    {\textbf{\textit{Interm.}}}   &    {\textbf{\textit{Hard}}} 
\\
\hline
\textbf{GPT-Neo}                    &       5.06  &	 4.27  &   2.06     &         19.62\%  &	18.52\%  &	 15.45\%        &          1.37\%  &  1.07\%  &	0.23\%       &	         0  &   0  &  0        \\

\rowcolor{gray!10} \textbf{CodeGen} &     7.63  &	 6.15  &   2.04     &         3.57\%  &	3.0\%  &	 1.48\%        &          1.14\%  &  0.84\%  &	0            &	         0  &   0  &  0        \\

\textbf{PolyCoder}                  &   4.52  &	 3.67  &   1.28         &         16.66\%  &	10.43\%  &	 -1.2\%       &          0.25\%  &  0.25\%  &	0.23\%        &	         0  &   0  &  0        \\

\rowcolor{gray!10} \textbf{CodeT5}  &    16.64  &	13.81  &   5.95     &         27.84\%  &	26.51\%  &	 22.78\%      &        1.26\%  & 1.73\%  &	3.04\%            &	      0.42\%  &   0.31\%  &  0    \\

\textbf{ChatGPT}                    &   12.81  &	11.8  &   \textbf{8.98}    &  \textbf{34.91\%}  &	30.65\%  &	 18.8\%    &   \textbf{88.96\%}  & \textbf{88.24\%}  &	\textbf{86.26\%}   &	     \textbf{49.53\%}  &   \textbf{44.08\%}  &  \textbf{28.95\%}    \\

\hdashline 

\rowcolor{gray!10} \textbf{E-code}  & \textbf{19.29} &	\textbf{17.1}  &   8.58    &   33.33\%  &	\textbf{32.67\%}  &	 \textbf{30.14\%}   &   82.27\%  & 79.35\%  &	67.93\%        &	   1.62\%  &   1.29\%  &  0    \\

\hline
\textbf{Average}                    &  10.99  &  9.47  &  4.82          &          22.66\%  &  20.3\%  &  14.58\%          &      29.21\%   &   28.58\%  &  26.28\%        &      8.6\%  &  7.6\%  &  4.83\%   \\
\hline
\end{tabular}
}
\label{tab:Difficulty}
\end{center}
\end{table}

% \begin{table}[htbp]
% \caption{Overall performance of decoder models CodeGen and PolyCoder on the ACEOB test set. (\%)}
% \begin{center}
% \begin{tabular}{@{}lcccccccccccc@{}}
% \hline

% \textit{\textbf{Models}}  &  \textit{\textbf{CodeBLEU}}      &   \textbf{\textit{EGR}}      &     \textbf{\textit{Compilation}}    &     \textbf{\textit{I/O}}    
%                           &  \textit{\textbf{CodeBLEU}}      &   \textbf{\textit{EGR}}      &     \textbf{\textit{Compilation}}    &     \textbf{\textit{I/O}}    
%                           &  \textit{\textbf{CodeBLEU}}      &   \textbf{\textit{EGR}}      &     \textbf{\textit{Compilation}}    &     \textbf{\textit{I/O}}    

% \\ \midrule

% \textbf{GPT-Neo}                           & 4.16  &	 9.08\%  &	1.25\%  &	0              \\ 

% \rowcolor{gray!10} \textbf{CodeGen}        & 4.68  &	 -3.97\%  &	0.4\%  &	0             \\

% \textbf{PolyCoder}                           & 2.67  &	11.18\%  &	0.36\%  &	0                 \\

% \rowcolor{gray!10} \textbf{CodeT5}           & 10.69  &	 6.44\%  &	1.69\%  &	0.13\%                        \\

% \hdashline 

% \textbf{ChatGPT}                              &    10.4  & 26.14\%  &	\textbf{87.16\%}  &	\textbf{37.21\%}                            \\

% \hline

% \rowcolor{gray!10} \textbf{E-code}          &  \textbf{14.44}  & \textbf{41.98\%}  &	76.94\%  &	0.93\%                \\ 

% \hline
% \end{tabular}
% \label{tab:BaseLine}
% \end{center}
% \end{table}

\subsection{RQ3: Efficiency-Driven Filtering Mechanism}\label{sec:7.3}

Limiting \textbf{Maximum Sampling} to \textbf{Five}. AlphaCode demonstrated the potential of improving model performance through large-scale sampling. ChatGPT, with its vast model parameters, also showcased impressive capabilities. However, both these methods come with noticeable computational overheads. In contrast, our small-scale sampling method leans more towards enhancing the computational efficiency of smaller models. Hence, to accommodate these small-scale model's computational capacities, we decided to limit the maximum sampling number to five.

Emphasis on an \textbf{efficiency-driven} code \textbf{filtering mechanism}. When a model produces multiple code samples yet only one code submission is permitted, an effective filtering mechanism becomes crucial. In efficient programming competitions, the efficiency of running the code is  more vital than its functionality. This is because the input of the EPC task comes with inefficient code, and producing another functionally correct but inefficient code is pointless. Conversely, even if a code has efficient algorithmic logic and can't compile, it might still serve as a reference. When sampling is more than one, we implement an efficiency-first filtering method. For instance, when the model samples five codes, we only submit the one with the lowest predicted running time by ExecTimePredictor. Notably, since ExecTimePredictor doesn't rely on real running time data, it is effectively applicable in practical scenarios.

Considering the impact of \textbf{varying code submission numbers}. Traditional programming competitions usually require participants to submit only one solution for each problem. To evaluate the performance limit of our efficiency-first filtering mechanism and the effect of increased code submissions, we particularly examine E-code's performance when submission counts are three and five.

The \textbf{filtering mechanism} substantially \textbf{elevates} E-code's \textbf{performance}. Table \ref{tab:Filtering} comprehensively displays how our filtering mechanism bolsters E-code's performance and considers the effect of different code submission numbers. Relatively speaking, sampling three times offers a better cost-effectiveness than sampling five times. The right half of Figure \ref{fig:Code_Token} exhibits the token length distribution of the code produced by E-code under the influence of the filtering mechanism, revealing almost no length change. This confirms that our filtering mechanism doesn't alter the code's length but filters based on the learned code efficiency knowledge. Performance-wise, the right halves of Figures \ref{fig:CodeBLEU} demonstrate how the filtering mechanism affects the CodeBLEU score distribution distribution of the code generated by E-code. The filtering mechanism slightly amplifies the CodeBLEU score.

\begin{table}[htbp]
\caption{An Investigation into the Performance Enhancement of E-code through Filtering Mechanisms and the Implications of Augmented Code Submission Frequency.}
\begin{center}
\begin{tabular}{@{}cc : ccccc@{}}
\hline

 \textit{\textbf{Sample Size}}   &   \textbf{\textit{Submission Count}}  
& \textit{\textbf{CodeBLEU}}  &   \textbf{\textit{EGR}}  &  \textbf{\textit{Compilation}}  &   \textbf{\textit{I/O}}  \\ 

\hline

\rowcolor{gray!10}           \textbf{1}     &     \textbf{1}     & 14.44  &	 30.74\%        &  	76.94\%    &	  0.93\%           \\

\hdashline 
  
                             \textbf{3}     &     \textbf{1}     & 14.7     &	49.42\%              &	68.69\%    &	0.63\%               \\

\rowcolor{gray!10}          \textbf{3}     &      \textbf{3}           & 15.54    & 54.98\%                &	88.99\%      &	0.97\%           \\
 
                             \textbf{5}     &     \textbf{1}     & 14.71	   & 54.98\%               &	70.06\%    &	0.63\%           \\

\rowcolor{gray!10}          \textbf{5}     &      \textbf{5}     & 15.83     & 59.37\%               &	92.56\%      &	1.04\%           \\

\hline
\end{tabular}
\label{tab:Filtering}
\end{center}
\end{table}

\subsection{RQ4: Analysis of Failure Causes}\label{sec:7.4}

\begin{figure}
    \centering
    \includegraphics[width=0.95\linewidth]{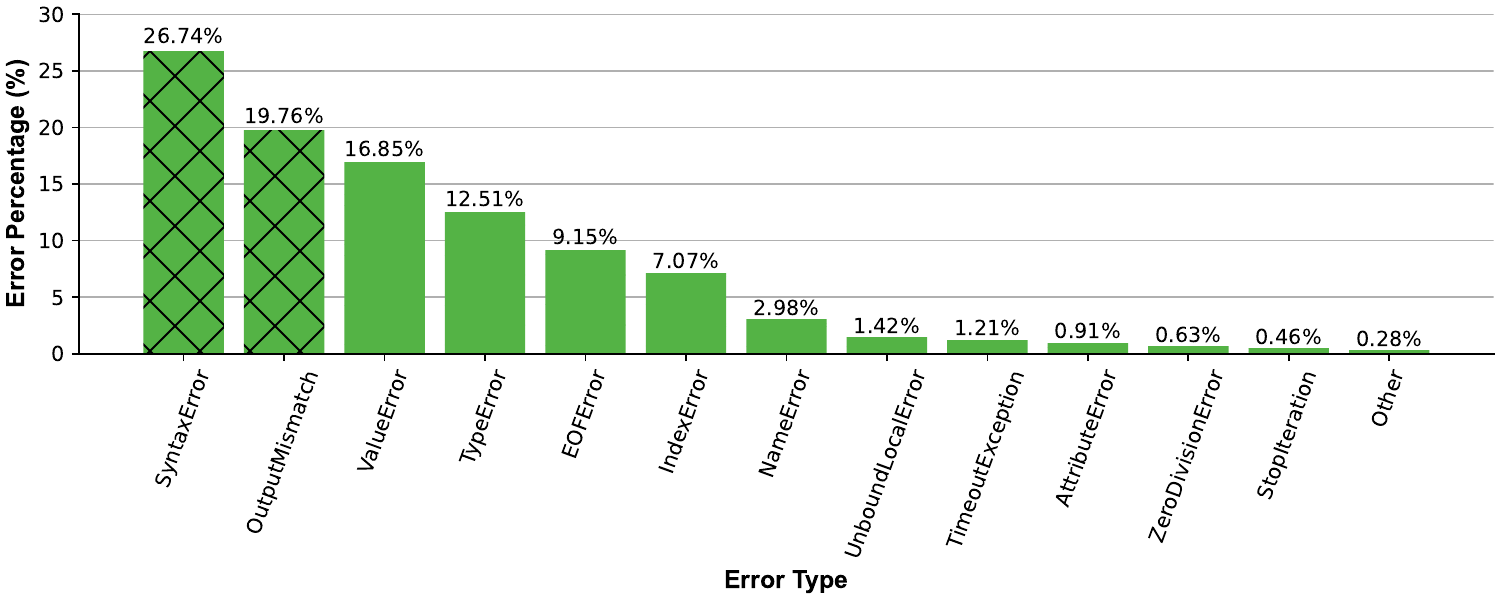}
    \caption{The primary error types encountered by E-code during EPC tasks.}
    \label{fig:ErrorType}
\end{figure}

To delve deeper into the challenges within the EPC task, we analyzed the \textbf{reasons} for E-code and ChatGPT's \textbf{failures} in this task. As depicted in Figure \ref{fig:ErrorType}, E-code's main error types when confronting the EPC task are enumerated. Among them, aside from ``SyntaxError'' being non-compilable, all other outcomes are classified as compilable. The ``OutputMismatch'' error occurs when the code doesn't pass all I/O tests, representing the second-largest proportion of errors in E-code. Correspondingly, ChatGPT's errors are quite concentrated, with ``OutputMismatch'' accounting for 63.91\% and ``SyntaxError'' constituting 17.97\%. These figures reflect that current large-scale language models, in the efficient code generation task, still struggle to flawlessly handle all I/O scenarios.

\subsection{RQ5: Ablation Analysis}\label{sec:7.5}

We report on the \textbf{ablation} studies of \textbf{various E-code components}. The experimental results, as presented in Table \ref{tab:Ablation}, lucidly depict the significance of each component within E-code. To verify the contribution of the expert encoder component, we specially trained an E-code version devoid of this expert encoder component, maintaining the same parameter count for both versions. In this control version, we used a complete Bert model to replace the expert encoder component. Notably, the encoder of the E-code without the expert component is consistent with traditional models in terms of feature extraction, i.e., it extracts overall features from input data. Furthermore, we also verified AST purification, uniform handling of variable and function names, and two model merging techniques to ascertain their indispensability.

\begin{table}[htbp]
\caption{Ablation experiments of components in E-code.}
\begin{center}
\begin{tabular}{@{}lcccc@{}}
\hline

\textit{\textbf{Models}}               & \textit{\textbf{CodeBLEU}}      &   \textbf{\textit{EGR}}      &     \textbf{\textit{Compilation}}    &     \textbf{\textit{I/O}}     \\ 

\hline

\textbf{No AST purification}                         &   5.44	        & -112.92\%          &	65.39\%       &	0.26\%                    \\

\rowcolor{gray!10} \textbf{No uniformity}            &  9.86	        & -105.02\%          &	65.75\%       &	0.26\%                        \\

\rowcolor{gray!10} \textbf{Using RELU}              &    11.93	        & 30.33\%         &	  64.41\%      &	0.45\%                                    \\

\rowcolor{gray!10} \textbf{No expert group E-code}   &   12.51          &	 28.25\%       &	71.70\%       &   	0          \\ 

\textbf{Uniform variable names}                      &  12.99	        & 24.2\%        &	74.6\%       &	0.56\%                              \\

\textbf{Using 8-head attention mechanism}           &   14.43   	     & 30.57\%            &	76.83\%        &	0.74\%                                  \\

\hline

\rowcolor{gray!10} \textbf{E-code}                  &  \textbf{14.44}	         & \textbf{30.74\%}           &	\textbf{76.94\%}          &	\textbf{0.93\%}                       \\ 

\hline
\end{tabular}
\label{tab:Ablation}
\end{center}
\end{table}

\section{Threats to Validity}\label{sec:8}

\subsection{Internal validity}\label{sec:8.1}

Our method has limitations in three aspects, and we analyze the potential causes.

\begin{itemize}
\item \textbf{Limitations of the expert encoder group.} The expert encoder group does not excel when dealing with tasks having only one type of input. However, when handling inputs that can be divided into multiple parts, this encoder group showcases unique advantages. Due to this design, E-code falls short in processing tasks with single input types. It's worth noting that, compared to competitive programming problems, the NL description of basic programming problems is usually shorter, reflecting more of a translation task. In this context, simply using this description as a code comment input to LMs means the advantage of the expert encoder group is not prominent in this aspect.
\item \textbf{ExecTimePredictor is biased.}  Since CodeExecTimeDB only includes functionally correct codes, ExecTimePredictor might exhibit biases when predicting non-compilable codes. Such biases could affect the accuracy of our code filtering method.
\item \textbf{E-code's functional accuracy is not on par with ChatGPT.} E-code seems unable to generate efficient code while maintaining functionality, even though it has a high compilation success rate. In its pursuit of code efficiency, E-code might excessively reduce code length, leading to functional losses. Moreover, given the granularity of the code, even a single token error could cause functional issues. This demands a high precision from the model. Considering the trade-off between model size and performance, our E-code model is only equipped with 401M parameters, which might limit its learning capability.
\end{itemize}

\subsection{External validity}\label{sec:8.2}

Currently, E-code focuses on optimizing single files, but optimization usually involves multiple classes or files. To extend to multi-file performance enhancement, we can look at expanding the context \cite{ewashemnlp} and input embedding matrix with expanded transformers \cite{Drain2021DeepDebugFP} and add more expert encoders in the encoder part of E-code to handle contexts from different classes or files. A key challenge is identifying which modifications are relevant; a potential solution is to construct method dependency information between files.

\section{Conclusion}\label{sec:9}

We introduced E-code, designed for generating efficient codes in the EPC task. E-code extracts features of different input types through various expert encoders and allocates different tasks to different components of the model. Based on rigorous experimental evaluations, E-code achieves a 54.98\% performance improvement, surpassing existing baseline models. We also conducted detailed ablation studies, confirming that all components of E-code are indispensable. E-code's expert encoder group offers a flexible encoder strategy, which can be expanded and improved in multiple ways to address various tasks.

% Use the style of numbering in square brackets.
% If nothing is used, default style will be taken.
%\begin{enumerate}[a)]
%\item 
%\item 
%\item 
%\end{enumerate}  

%\begin{itemize}
%\item 
%\item 
%\item 
%\end{itemize}  

%\begin{description}
%\item[]
%\item[] 
%\item[] 
%\end{description}  

% % Figure
% \begin{figure}[<options>]
% 	\centering
% 		\includegraphics[<options>]{}
% 	  \caption{}\label{fig1}
% \end{figure}

% \begin{table}[<options>]
% \begin{tabular*}{\tblwidth}{@{}LL@{}}
% \toprule
% \midrule
%  & \\
%  & \\
%  & \\
%  & \\
% \bottomrule
% \end{tabular*}
% \end{table}

% Uncomment and use as the case may be
%\begin{theorem} 
%\end{theorem}

% Uncomment and use as the case may be
%\begin{lemma} 
%\end{lemma}

%% 附录 The Appendices part is started with the command \appendix;
% \section{}\label{}


\bibliographystyle{elsarticle-num}
% \bibliographystyle{cas-model2-names}
% \bibliography{ref}

% % Loading bibliography database
% \bibliography{}

% % Biography
% % Here goes the biography details.

% % Here goes the biography details.

\end{document}